\documentclass[conference]{IEEEtran}
\IEEEoverridecommandlockouts

\usepackage{cite}
\usepackage{amsmath,amssymb,amsfonts}
\usepackage{algorithmic}
\usepackage{graphicx}
\usepackage{textcomp}
\usepackage{xcolor}
\usepackage{url}
\usepackage{eurosym}

\def\BibTeX{{\rm B\kern-.05em{\sc i\kern-.025em b}\kern-.08emT\kern-.1667em\lower.7ex\hbox{E}\kern-.125emX}}

\usepackage[top=0.75in,bottom=1in,left=0.68in,right=0.68in]{geometry}

\newcommand{\user}{{\sf User}}
\newcommand{\CS}{{\sf CS}}
\newcommand{\SP}{{\sf SP}}
\newcommand{\TI}{{\sf TI}}

\begin{document}
\title{An End-To-End Encrypted Cache System with\\ Time-Dependent Access Control\thanks{$\ast$~An extended abstract appeared at the 9th International Conference on Information Systems Security and Privacy, ICISSP 2023~\cite{EY23}. This is the full version. }}

\date{\today}

\author{\IEEEauthorblockN{Keita Emura}
\IEEEauthorblockA{\textit{National Institute of}\\\textit{Information and}\\\textit{Communications Technology}\\\textit{Japan}}\
\and
\IEEEauthorblockN{Masato Yoshimi}
\IEEEauthorblockA{\textit{TIS Inc., Japan}}\
}

\maketitle

\begin{abstract}
Due to the increasing use of encrypted communication, such as Transport Layer Security (TLS), encrypted cache systems are a promising approach for providing communication efficiency and privacy. Cache-22 is an encrypted cache system  (Emura et al. ISITA 2020) that makes it possible to significantly reduce communication between a cache server and a service provider. 
In the final procedure of Cache-22, the service provider sends the corresponding decryption key to the user via TLS and this procedure allows the service provider to control which users can access the contents. For example, if a user has downloaded ciphertexts of several episodes of a show, the service provider can decide to provide some of the contents (e.g., the first episode) available for free while requiring a fee for the remaining contents. However, no concrete access control method has been implemented in the original Cache-22 system. 
In this paper, we add a scalable access control protocol to Cache-22. Specifically, we propose a time-dependent access control that requires a communication cost of $O(\log T_{\sf max})$ where $T_{\sf max}$ is the maximum time period. 
Although the protocol is stateful, we can provide time-dependent access control with scalability at the expense of this key management. 
We present experimental results and demonstrate that the modified system is effective for controlling access rights. 
We also observe a relationship between cache capacity and network traffic because the number of duplicated contents is higher than that in the original Cache-22 system, due to time-dependent access control. 
\end{abstract}

\begin{figure*}[t]
\centering
  \begin{minipage}[b]{0.45\linewidth}
    \centering
    ~\hspace{-5mm}\includegraphics[keepaspectratio, scale=0.5]{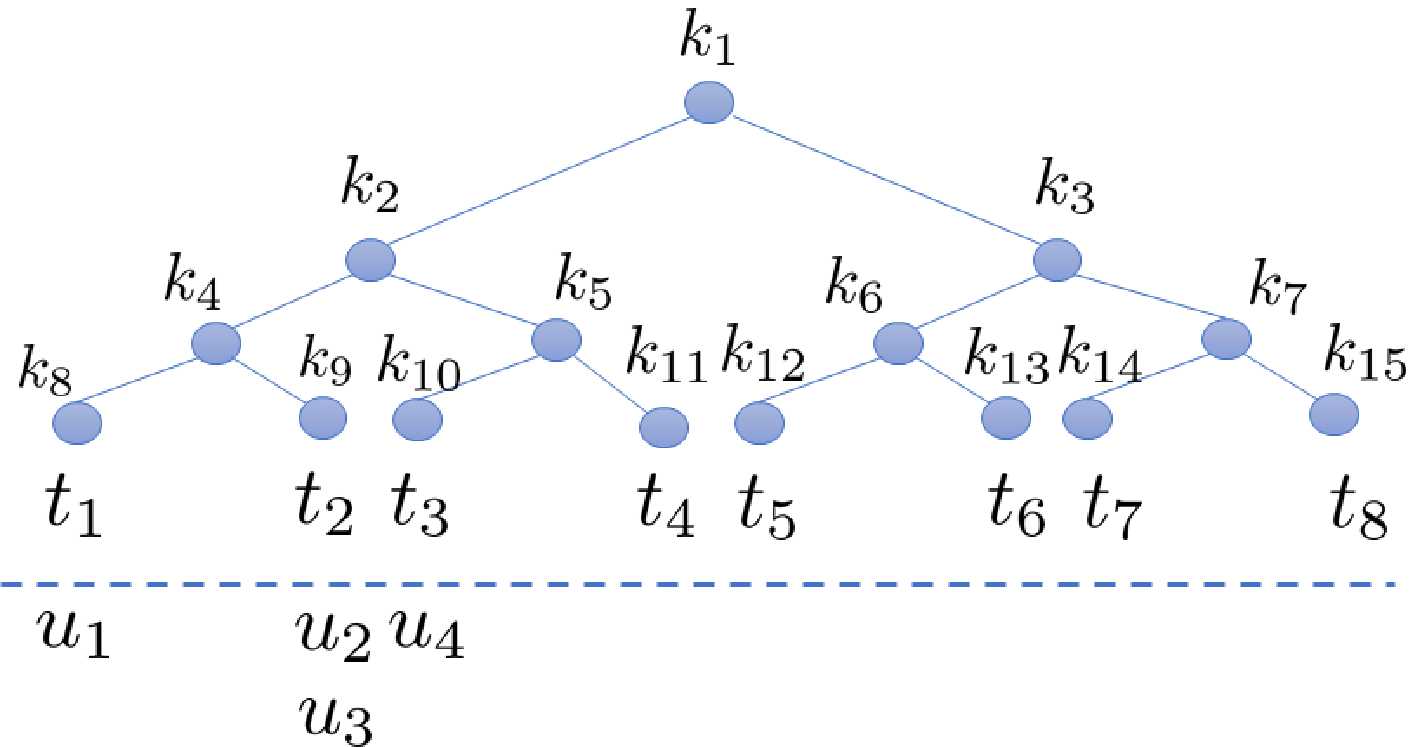}
    \caption{Time Tree ($T_{\sf max}=8$)}\label{tree1}
  \end{minipage}
  \begin{minipage}[b]{0.45\linewidth}
    \centering
    ~\hspace{5mm}\includegraphics[keepaspectratio, scale=0.5]{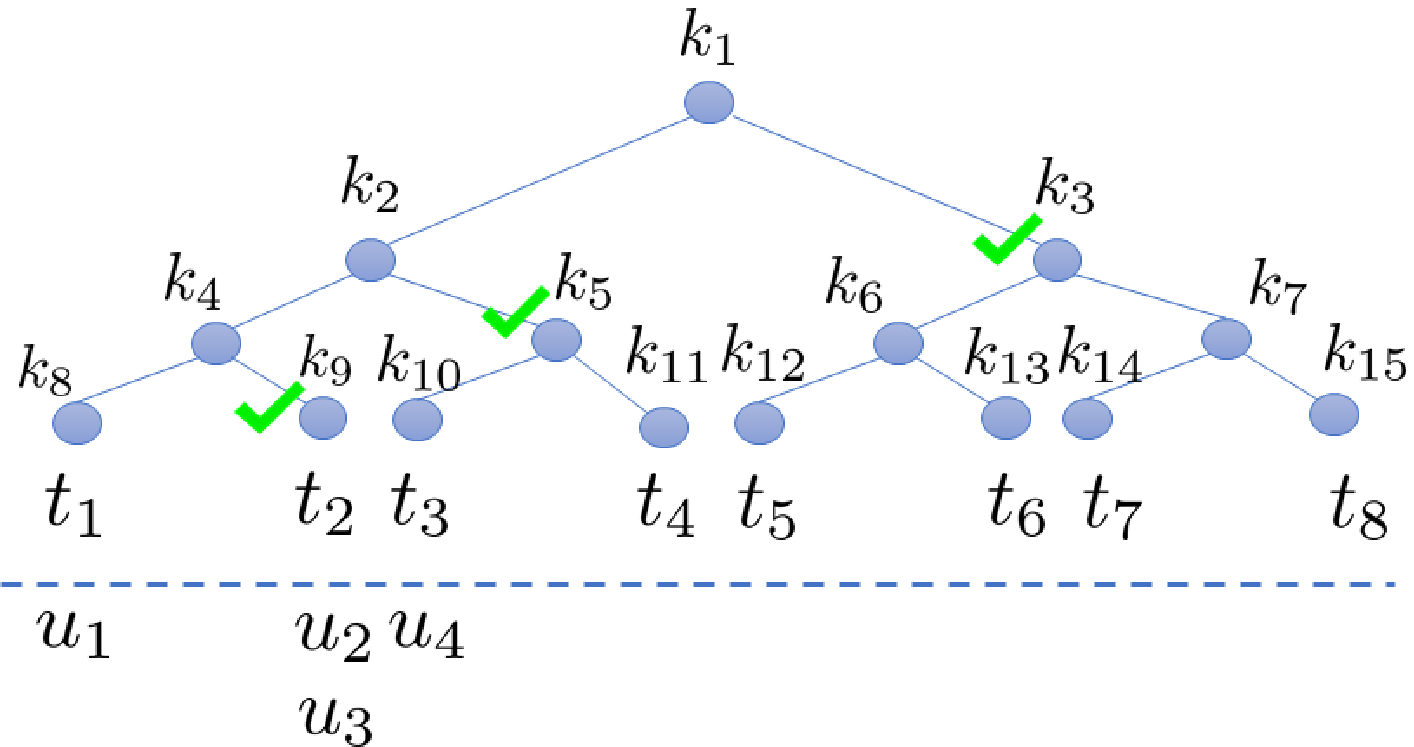}
    \caption{Time Tree (at time $t_2$)}\label{tree2}
  \end{minipage}
\end{figure*}

\section{Introduction}
\label{sec:introduction}

Cache systems are vital to reduce communication overhead on the Internet. However, it is not trivial to provide cache systems over encrypted communications because a cache server (CS) must verify whether it has a copy of a particular encrypted content, although information about the content is not revealed due to encryption. Thus, due to the increasing use of encrypted communication, such as Transport Layer Security (TLS), encrypted cache systems are a promising approach for providing communication efficiency and privacy. 

Leguay et al.~\cite{[LeguayPQS17]} proposed an encrypted cache system called CryptoCache. Although the contents are encrypted, CryptoCache allows users requesting the same content to be linked. Thus, Leguay et al. proposed an extension that prevents this linkability by employing a public key encryption (PKE) scheme. Emura et al.~\cite{EmuraMNY20,EmuraMNY22} further extended CryptoCache by proposing an encrypted cache system called Cache-22. The Cache-22 system not only provides unlinkability without employing PKE, but also presents a formal security definition in a cryptographic manner. 

\begin{figure*}[t]
\centering
    \includegraphics[keepaspectratio, scale=0.5]{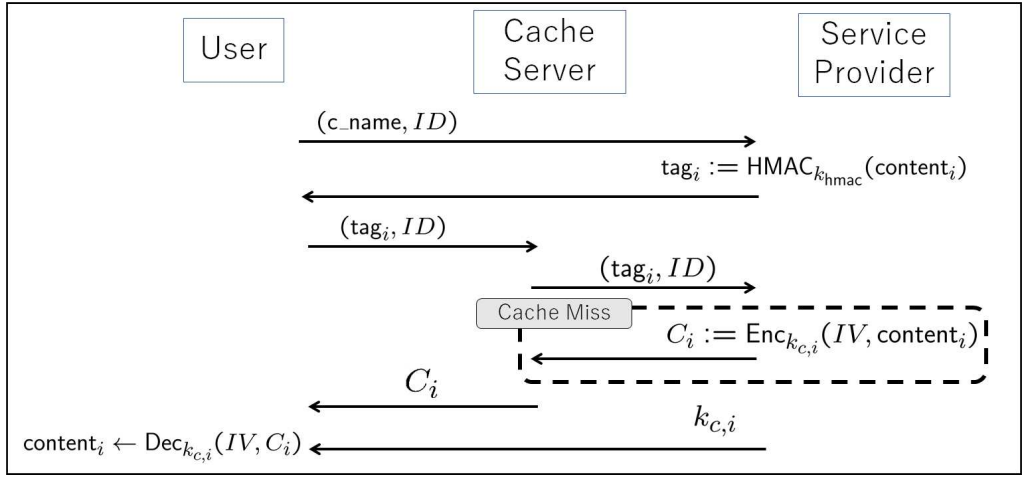}
    \caption{Cache-22 System~\cite{EmuraMNY20,EmuraMNY22}}\label{cache-22}
\end{figure*}

The Cache-22 system is briefly explained as follows and illustrated in Figure~\ref{cache-22} in Section~\ref{subsec_cache-22}. 
It is assumed that all communications are protected by TLS. A tag is assigned to each content, and it is assumed that no information about the content is revealed by the tag (e.g., it can be generated using hash-based message authentication code (HMAC), because it is a pseudorandom function~\cite{Bellare15}). 
The service provider (SP) encrypts content and stores the ciphertext and corresponding tag on a CS. When a user requests the content, the user sends a request to the SP. Then, the SP sends the corresponding tag back to the user. The user then sends the tag to the CS. If the tag is stored on the CS, the CS sends the corresponding ciphertext to the user and the user information to the SP. Finally, the SP sends the corresponding decryption key to the user. Because the size of the tag is much smaller than the size of the content (ciphertext), the Cache-22 system makes it possible to significantly reduce communications between a CS and the SP. 
Because the Cache-22 system can employ any cipher suite, seven cipher suites, including National Institute of Standards and Technology (NIST) Post-Quantum Cryptography (PQC) candidates~\cite{BIKE,BosDKLLSSSS18,NTRU,SABER} are employed. 

\medskip\noindent\textbf{Adding Access Control to Cache-22}: 
In the final procedure of the Cache-22 system, the SP sends the corresponding decryption key to the user. Emura et al.~\cite{EmuraMNY20,EmuraMNY22} claimed that this procedure allows the SP to control which users can access the contents. For example, if a user has downloaded ciphertexts of several episodes of a show, the SP can allow some of the contents (e.g., the first episode) to be available for free while requiring a fee for the remaining contents. However, the authors did not provide a concrete access control method. 

A naive solution is to add an authentication protocol, such as classical ID/password authentication, before the SP sends the corresponding decryption key to the user. This method is effective; however, it is not scalable. That is, the SP must send the decryption key individually for $N$ users, which leads to a communication cost of $O(N)$. 

\subsection{Our Contribution}

In this paper, we add a scalable access control protocol to the Cache-22 system. Specifically, we propose time-dependent access control, which requires a communication cost of $O(\log T_{\sf max})$ using the Naor--Naor--Lotspiech (NNL) framework~\cite{NaorNL01} where $T_{\sf max}$ is the maximum time period. In the original NNL framework, each user is assigned to a leaf node of a binary tree which provides broadcast encryption in which the encryptor specifies who can decrypt the ciphertext. In our proposed protocol, each time period is assigned to a leaf node (multiple users are assigned to the same node if they have the same access rights). 
Briefly, let $\TI=[1,T_{\sf max}]$ be a time interval where $T_{\sf max}\in\mathbb{N}$ and assume that $T_{\sf max}=2^m$ for some $m\in\mathbb{N}$. 
Then, each time $t\in\TI$ is assigned to a leaf node of a binary tree that has $2^m$ leaves. This time period indicates how long the content is available. For example, $t$ can represent a day, a week, a month, and so on. The SP encrypts each content according to the time it is available. 
This NNL-based time-dependent control technique has been employed in other cryptographic primitives, such as attribute-based encryption for range attributes~\cite{AttrapadungHOOW16} and group signatures with time-bound keys~\cite{EmuraHI20}. However, to the best of our knowledge, no encrypted cache system with this technique has been proposed so far. 

\medskip\noindent\textbf{Toy Example}. 
Figure~\ref{tree1} presents an example when $T_{\sf max}=8$. A key is assigned to each node (from $k_1$ to $k_{15}$), and each user is also assigned to a leaf according to the corresponding access rights. In this example, there are four users, $u_1$, $u_2$, $u_3$, and $u_4$. Here, $u_2$ and $u_3$ are assigned to the same leaf indicating that they have the same access rights. 
Each user has keys on the path, namely, from their own leaf node to the root ($u_2$ and $u_3 $ have keys $k_1$, $k_2$, $k_4$, and $k_9$). 
At time $t_1$, the SP encrypts the content \lq\lq Episode 1" using $k_1$. Now, all users can obtain the content because they have $k_1$. At time $t_2$, the SP encrypts the content \lq\lq Episode 2" using $k_3$, $k_5$, and $k_9$ (see Figure~\ref{tree2}); that is, there are three ciphertexts. In this case, $u_1$ has no rights to obtain the content whereas the other users still have at least one decryption key. Similarly, $k_3$ and $k_5$ are used for encryption at time $t_3$, and $k_3$ and $k_{11}$ are used for encryption at time $t_4$. 
In this method, the number of ciphertexts is $O(\log T_{\sf max})$, and each user manages $O(\log T_{\sf max})$-size decryption keys. 
One shortcoming of this construction based on the NNL framework compared to the original Cache-22 is that each user must manage the decryption keys; that is, the protocol is stateful. Nevertheless, this construction provides time-dependent access control with scalability at the expense of key management. 

\section{Preliminaries}

\subsection{Cache-22 System}
\label{subsec_cache-22}

In this section, we introduce the Cache-22 system. 
 A tag is assigned to each content, and it is assumed that no information 
about the content is revealed by the tag. The SP encrypts the content and stores a tag and ciphertext pair on the CS.
In the implementation proposed by~\cite{EmuraMNY20,EmuraMNY22}, there are multiple CSs due to the color-based cooperative cache system~\cite{NakajimaYWY17}. For the sake of simplicity, we consider the case of a single CS. We assume that all communications between a user, CS, and SP are encrypted with TLS. 
Let $({\sf Enc},{\sf Dec})$ be a IND-CPA secure SKE scheme, where for a key $k\in\mathcal{K}$ and a message $M\in\mathcal{M}$, ${\sf Dec}_k(C)=M$ holds, where $C\leftarrow{\sf Enc}_{k}(M)$, $\mathcal{K}$ is the key space, and $\mathcal{M}$ is the message space. Here, IND-CPA stands for indistinguishability under chosen-plaintext attack. 
The upper-order 128 bits of tag are used as the initial vector ($IV$) for {\sf AES-GCM}~\cite{IwataS17}. Then, $IV$ is not reused for other encryption since the tag is pseudorandom. 
Let ${\sf CacheTbl}$ be the cache table managed by the CS which has the structure ${\sf CacheTbl}=\{({\sf tag}_i,C_i)\}$, and is initiated as $\emptyset$. Although we simply denote ${\sf CacheTbl}=\{({\sf tag}_i,C_i)\}$ here, we can employ any cache system. We also assume that a user knows the content name ${\sf c\_name}$, and that the SP can decide the corresponding content ${\sf content}_i\in\mathcal{M}$ from ${\sf c\_name}$. The flow of the Cache-22 system is illustrated in Figure~\ref{cache-22}, and the formal description of the system is provided as follows. 
The Cache-22 system consists of $({\sf GenTable},\allowbreak {\sf ContentRequest},\allowbreak {\sf SendContent},\allowbreak {\sf CacheRequest},\allowbreak {\sf SendKey},\allowbreak {\sf ObtainContent})$. 
 It should be noted that the SP sends the corresponding decryption key to a user via the ${\sf SendKey}$ algorithm. Because the SP needs to know the destination, each user sends own identity $ID$ to the CP in the ${\sf SendContent}$ protocol. 

\begin{itemize}
\item ${\sf GenTable}(1^\kappa,1^\lambda,{\sf SetOfContents})$: The table generation algorithm (run by the SP) takes as input security parameters $\kappa,\lambda\in\mathbb{N}$ and a set of contents ${\sf SetOfContents}\allowbreak=\{{\sf content}_i\}_{i=1}^n$. 
Randomly choose $k_{c,i}\leftarrow \mathcal{K}$ and compute ${\sf tag}_i\leftarrow {\sf HMAC}_{k_{\sf hmac}}({\sf content}_i)$ for each ${\sf content}_i\in\mathcal{M}$. 
Retrieve $IV$ from ${\sf tag}_i$, and encrypt ${\sf content}_i$ such that $C_i\leftarrow {\sf Enc}_{k_{c,i}}(IV,{\sf content}_i)$. Output a table ${\sf ConTbl}=\{({\sf content}_i,\allowbreak{\sf tag}_i,C_i,k_{c,i})\}$.

\smallskip
\item ${\sf ContentRequest}(\user({\sf c\_name},ID),\SP({\sf ConTbl}))$: The ${\sf ContentRequest}$ protocol between a user and the SP takes as input a content name ${\sf c\_name}$ and the user identity $ID$ from the user, and takes as input ${\sf ConTbl}$ from the SP. 

\begin{enumerate}
\item The user sends $({\sf c\_name},\allowbreak ID)$ to the SP via a secure channel. 

\item The SP decides ${\sf content}_i$ from ${\sf c\_name}$, and retrieves the corresponding $({\sf content}_i,{\sf tag}_i,C_i,k_{c,i})$ from ${\sf ConTbl}$. 

\item The SP sends ${\sf tag}_i$ to the user via the secure channel. 
\end{enumerate}

\smallskip
\item ${\sf SendContent}(\user({\sf tag}_i,ID),\CS({\sf CacheTbl}))$: The content sending protocol between a user and the CS takes as input $({\sf tag}_i,ID)$ from the user, and takes as input ${\sf CacheTbl}$ from the CS. 

\begin{enumerate}
\item The user sends a request $({\sf tag}_i,ID)$ to the CS via a secure channel. 

\item The CS checks whether ${\sf tag}_i$ is stored in ${\sf CacheTbl}$. 

\begin{itemize}
\item If yes, the CS retrieves $({\sf tag}_i,C_i)$ from ${\sf CacheTbl}$ by using ${\sf tag}_i$, sends $C_i$ to the user via the secure channel, and sends $({\sf tag}_i,ID)$ to the SP via the secure channel. 

\item If no, the CS runs the ${\sf CacheRequest}$ protocol with the SP (which is defined later), obtains $C_i$, stores $({\sf tag}_i,C_i)$ to ${\sf CacheTbl}$, and sends $C_i$ to the user via the secure channel. 
\end{itemize}
\end{enumerate}

\smallskip
\item ${\sf CacheRequest}(\CS({\sf tag}_i,ID),\SP({\sf ConTbl}))$: The cache request protocol between the CS and the SP takes as input $({\sf tag}_i,ID)$ from the CS, and takes as input ${\sf ConTbl}$ from the SP. 

\begin{enumerate}
\item The CS sends $({\sf tag}_i,ID)$ to the SP via the secure channel. 

\item The SP retrieves $({\sf content}_i,{\sf tag}_i,C_i,k_{c,i})$ from ${\sf ConTbl}$ by using ${\sf tag}_i$, and sends $C_i$ to the CS via the secure channel. 
\end{enumerate}

\smallskip
\item ${\sf SendKey}(ID,k_{c,i})$: The key sending algorithm run by the SP takes as input $(ID,k_{c,i})$. Send $k_{c,i}$ to the user whose identity is $ID$ via the secure channel. 

\smallskip
\item ${\sf ObtainContent}({\sf tag}_i,C_i,k_{c,i}))$: The content obtaining algorithm run by a user takes as input $({\sf tag}_i, C_i,k_{c,i})$. 
Retrieve $IV$ from ${\sf tag}_i$. 
Output ${\sf content}_i\leftarrow {\sf Dec}_{k_{c,i}}(IV,\allowbreak C_i)$. 
\end{itemize}

As mentioned in the introduction, there is room for adding an access control system before running the ${\sf SendKey}$ algorithm. 

\begin{figure*}[t]
\centering
    \includegraphics[keepaspectratio, scale=0.5]{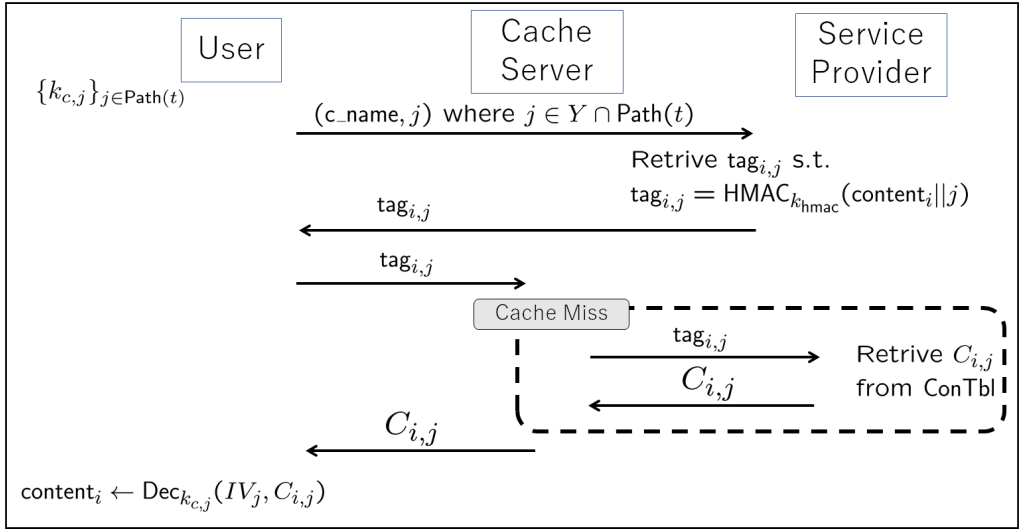}
    \caption{Cache-22 System with Time-Dependent Access Control}\label{cache-22-TDAC}
\end{figure*}

\subsection{NNL Framework}

In this section, we introduce the NNL framework which is called the complete subtree method. Let ${\sf BT}$ be a binary tree with $N$ leaves. For a leaf node $i$, let ${\sf Path}(i)$ be the set of nodes from the leaf to the root. Let ${\sf RSet}$ be the set of revoked leaves. For non leaf node $x$, let $x_{\sf left}$ be the left child of $x$ and $x_{\sf right}$ be the right child of $x$. 

\begin{enumerate}
\item Initialize $X,Y\leftarrow\emptyset$. 

\item For all $i\in {\sf RSet}$, add ${\sf Path}(i)$ to $X$. 

\item For all $x\in X$, if $x_{\sf left}\not\in X$ then add $x_{\sf left}$ to $Y$. If $x_{\sf right}\not\in X$ then add $x_{\sf right}$ to $Y$. 

\item If $|{\sf Rset}|=0$ then add the root node to $Y$. 

\item Output $Y$. 
\end{enumerate}

\noindent 
We denote $Y\leftarrow {\sf CompSubTree}({\sf BT},{\sf RSet})$. 
In the proposed time-dependent access control, a time period is assigned to a leaf, although each user is assigned to a leaf node in the original complete subtree method. Moreover, each leaf is sequentially revoked from the leftmost node. Then, the size of $Y$ is estimated as $|Y|=O(\log N)$ where $N:=T_{\sf max}$ in our protocol, which is scalable regardless of the number of revoked users in the system. 


\section{Cache System with Time-Dependent Access Control}
\label{sec:tdac}

In this section, we present our proposed protocol with time-dependent access control. Each content is encrypted with a time period $t$, and if a user is assigned to a time period $t^\prime$, then that user is allowed to obtain contents encrypted with $t$, where $t\leq t^\prime$. 
For the sake of simplicity, we assume that the access rights of all users are determined in advance. 
As a remark, we may be able to assume that all contents are encrypted and the SP stores all ciphertexts to the CS regardless of whether they are requested by a user or not. Then, a request sent by a user will always be successful (cache hits). However, this situation is unrealistic because the storage size of the CS will drastically increase. Thus, the SP adds new contents after receiving a user request. 

Let $T_{\sf max}$ be the maximum time period where $T_{\sf max}\in\mathbb{N}$ and assume that $T_{\sf max}=2^m$ for some $m\in\mathbb{N}$. Each time period $t\in \TI=[1,T_{\sf max}]$ is assigned to a leaf node. If a user is assigned to a time period $t$, ${\sf Path}(t)$ denotes the set of nodes from the leaf node (which is assigned to $t$) to the root node. 
Let ${\sf CacheTbl}$ be initialized as $\emptyset$. 
In the original Cache-22 system, each tag is generated by the corresponding content such as ${\sf tag}_i\leftarrow {\sf HMAC}_{k_{\sf hmac}}({\sf content}_i)$. In our proposed system, one content is multiply encrypted due to the NNL framework. To clarify which ciphertext should be sent to a user, each tag is generated by both the corresponding content and the corresponding index (determined by the NNL framework) such as ${\sf tag}_{i,j}\leftarrow {\sf HMAC}_{k_{\sf hmac}}({\sf content}_i||j)$. 

The proposed Cache-22 system with time-dependent access control consists of $({\sf KeyGen},\allowbreak {\sf SendKey},{\sf GenTable},\allowbreak {\sf ContentRequest},\allowbreak {\sf CacheRequest}, \allowbreak {\sf SendContent},\allowbreak {\sf ObtainContent})$ as illustrated in Figure~\ref{cache-22-TDAC}. Unlike to the original Cache-22 system, in the proposed system, all keys are generated in advance, i.e., they are independent of the contents. Thus, we add the ${\sf KeyGen}$ algorithm. 
Moreover, for a user with identity $ID$, the SP sends keys in accordance with the user's access rights. 
Thus, we run the ${\sf SendKey}$ algorithm before the ${\sf GenTable}$ algorithm. 

\begin{itemize}

\item ${\sf KeyGen}(1^m)$: The key generation algorithm takes as a security parameter $m\in\mathbb{N}$. For $j=1,2,\ldots,2^{m+1}-1$, randomly choose $k_{c,j}\leftarrow \mathcal{K}$ and output $\{k_{c,j}\}_{j=1}^{2^{m+1}-1}$.

\item ${\sf SendKey}(ID,t,\allowbreak \{k_{c,j}\}_{j=1}^{2^{m+1}+1})$: The key sending algorithm run by the SP takes as input $(ID,t,\allowbreak \{k_{c,j}\}_{j=1}^{2^{m+1}+1})$. For all $j\in {\sf Path}(t)$, send $k_{c,j}$ to the user with identity $ID$ via a secure channel. 

\item ${\sf GenTable}(1^\kappa,1^\lambda,\{k_{c,j}\}_{j=1}^{2^{m+1}+1},{\sf SetOfContents})$: The table generation algorithm (run by the SP) takes as input security parameters $\kappa,\lambda\in\mathbb{N}$, a set of keys $\{k_{c,j}\}_{j=1}^{2^{m+1}+1}$, and a set of contents ${\sf SetOfContents}\allowbreak=\{{\sf content}_i\}_{i=1}^n$. 
 For $i=1,2,\ldots,n$, let $t_i\in [1,T_{\sf max}]$ be the time period of ${\sf content}_i$. For all $j\in {\sf Path}(t_i)$, compute ${\sf tag}_{i,j}\leftarrow {\sf HMAC}_{k_{\sf hmac}}({\sf content}_i||j)$, retrieve $IV_j$ from ${\sf tag}_{i,j}$, and encrypt ${\sf content}_i$ such that $C_{i,j}\leftarrow {\sf Enc}_{k_{c,j}}(IV_j,{\sf content}_i)$. Output a table ${\sf ConTbl}=\{({\sf content}_i,\allowbreak\{({\sf tag}_{i,j},C_{i,j},k_{c,j})\}_{j\in {\sf Path}(t_i)})\}$. 

\item ${\sf ContentRequest}(\user({\sf c\_name},t,t_{\sf curr}),\SP({\sf ConTbl}))$: The ${\sf ContentRequest}$ protocol between a user and the SP takes as input a content name ${\sf c\_name}$, the time period of the user $t$, and the current time period $t_{\sf curr}$ from the user, and takes as input ${\sf ConTbl}$ from the SP. 

\begin{table*}[t]
  \caption{Libraries included in the modules.}
  \label{tab:Library}
  \centering
  \begin{tabular}{lll}\\\hline\hline
    & Version & Description \\\hline\hline
    Go & go1.18.6-devel-cf & Custom Go language~\cite{GowithCIRCL}\\\hline
    CIRCL & v1.2.0 & Collection of PQC primitives\\\hline
    labstack/echo & v4.9.0 & WebAPI Framework \\\hline
    syndtr/goleveldb & v1.0.0 & Non-volatile key-value store to configure LRU cache\\\hline
    math/rand & Standard & Zipf function to generate content requests by user \\\hline\hline
  \end{tabular}
\end{table*}

\begin{table*}[t]
  \caption{Host configuration}
  \label{tab:HostConfiguration}
  \centering
  \centering
  \begin{tabular}{lrl}\hline\hline
    & Specifications & Description  \\\hline\hline
    Instance type & c5.4xlarge & up to 0.856 [USD/hour] \\\hline
    vCPU  [Core] & 16 & Intel Xeon Platinum 8275CL @ 3.00GHz \\\hline
    Memory [GiB] & 32 & \\\hline
    Network [Gbps] & up to 10 & \\\hline
    Operating system & Amazon Linux 2 & Kernel 5.10.135-122.509 \\\hline
    Number of hosts & 3 & for CS, SP, and User \\\hline\hline
  \end{tabular}
\end{table*}

\begin{enumerate}

\item The user runs $Y\leftarrow {\sf CompSubTree}({\sf BT},[1,t_{\sf curr}-1])$ where ${\sf BT}$ is a binary tree with $2^m$ leaves. If $Y \cap {\sf Path}(t)=\emptyset$, then abort. 

\item The user chooses $j\in Y \cap {\sf Path}(t)$. 

\item The user sends $({\sf c\_name},j)$ to the SP via a secure channel. 

\item The SP decides ${\sf content}_i$ from ${\sf c\_name}$ and retrieves the corresponding $({\sf tag}_{i,j},C_{i,j})$ from ${\sf ConTbl}$ where ${\sf tag}_{i,j}\leftarrow {\sf HMAC}_{k_{\sf hmac}}({\sf content}_i||j)$. 

\item The SP sends ${\sf tag}_{i,j}$ to the user via the secure channel. If there is no such entry, then return error. 
\end{enumerate}

\smallskip
\item ${\sf SendContent}(\user({\sf tag}_{i,j}),\CS({\sf CacheTbl}))$: The content sending protocol between a user and the CS takes as input ${\sf tag}_{i,j}$ from the user, and takes as input ${\sf CacheTbl}$ from the CS. 

\begin{enumerate}
\item The user sends a request ${\sf tag}_{i,j}$ to the CS via a secure channel. 
\item The CS checks whether ${\sf tag}_{i,j}$ is stored on ${\sf CacheTbl}$. 

\begin{itemize}
\item If yes, the CS retrieves $({\sf tag}_{i,j},C_{i,j})$ from ${\sf CacheTbl}$ by using ${\sf tag}_{i,j}$, sends $C_{i,j}$ to the user via the secure channel. 

\item If no, the CS runs the ${\sf CacheRequest}$ protocol with the SP (which is defined later), obtains $C_{i,j}$, stores $({\sf tag}_{i,j},C_{i,j})$ to ${\sf CacheTbl}$, and sends $C_{i,j}$ to the user via the secure channel. 
\end{itemize}

\end{enumerate}

\smallskip
\item ${\sf CacheRequest}(\CS({\sf tag}_{i,j}),\SP({\sf ConTbl}))$: The cache request protocol between the CS and the SP takes as input ${\sf tag}_{i,j}$ from the CS, and takes as input ${\sf ConTbl}$ from the SP. 

\begin{enumerate}
\item The CS sends ${\sf tag}_{i,j}$ to the SP via the secure channel. 

\item The SP retrieves the corresponding $({\sf tag}_{i,j},C_{i,j})$ from ${\sf ConTbl}$ by using ${\sf tag}_{i,j}$, and sends $C_{i,j}$ to the CS via the secure channel. 
\end{enumerate}

\smallskip
\item ${\sf ObtainContent}({\sf tag}_{i,j},C_{i,j},k_{c,j}))$: The content obtaining algorithm run by a user takes as input $({\sf tag}_{i,j},C_{i,j},k_{c,j})$. Retrieve $IV_j$ from ${\sf tag}_{i,j}$. 
Output ${\sf content}_i\leftarrow {\sf Dec}_{k_{c,j}}(IV_j,\allowbreak C_{i,j})$. 

\end{itemize}

As a side effect, users do not need to send their identity to the CS in the proposed system. In contrast, in the original Cache-22 system, users must send their identity to the CS because the SP must send the corresponding decryption key to the user, and the CS thus needs to forward the identity to the SP to provide the destination. The proposed system can thus help hide the user's identity from the CS and preserve privacy. 

\section{Implementation and Results}


\subsection{Cipher Suite}
First, we decide the underlying cipher suite as 

\begin{itemize}
\item \small{\texttt{TLS\_Kyber\_ECDSA\_WITH\_AES\_256\_GCM\_SHA256}}
\end{itemize}

\noindent 
We employed Kyber (Crystals-Kyber)~\cite{BosDKLLSSSS18} which was selected for NIST PQC standardization in July 2022. Kyber (Crystals-Kyber) is a lattice-based scheme and is secure under the MLWE assumption where MLWE stands for the module learning with errors. In our implementation, we employed Kyber512 to provide 128-bit security. Specifically, we installed the X25519Kyber512Draft00 key agreement in our experiment. 
As in the original Cache-22 system, the proposed system can employ other PQC such as BIKE~\cite{BIKE}, NTRU~\cite{NTRU}, and SABER~\cite{SABER}. 

We also considered the underlying SKE scheme and hash function to be secure against the Grover algorithm~\cite{Grover98}, we expanded the key length twice and employed AES256 (specifically, AES-GCM-256) and SHA256. 
As a remark, as in the original Cache-22 implementation, we did not consider post-quantum authentication.%
\footnote{We refer the comment by Alkim et al.~\cite{AlkimDPS16}, \lq\lq \emph{the protection of stored transcripts against future decryption using quantum computers is much more urgent than post-quantum authentication. Authenticity will most likely be achievable in the foreseeable future using proven pre-quantum signatures and attacks on the signature will not compromise previous communication".}} 

\begin{figure*}[t]
\centering
\includegraphics[width=1.0\linewidth]{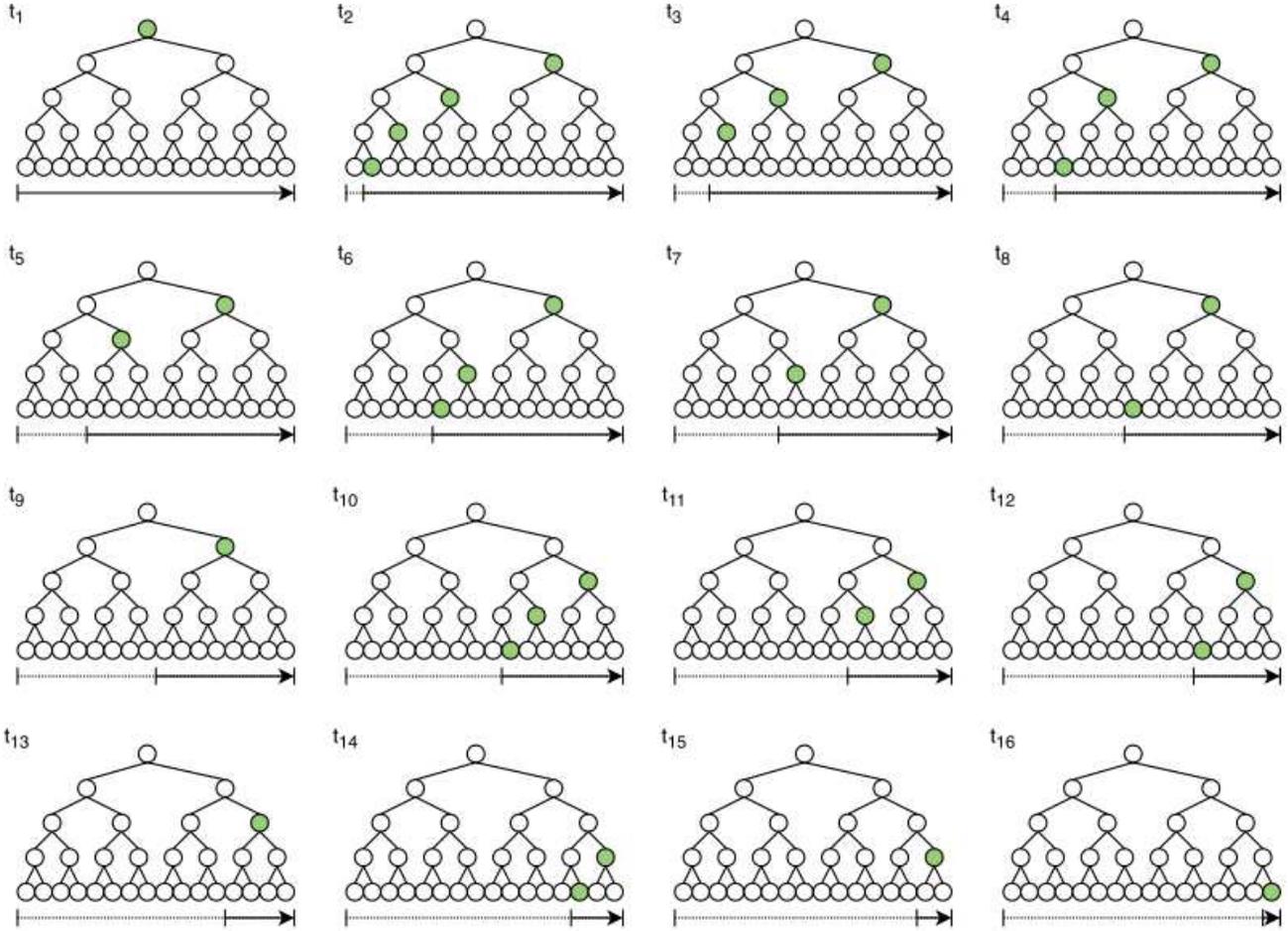}
\caption{Experimental Flow}\label{fig:experimental}
\end{figure*}

\begin{table*}[t]
  \caption{Experimental Setup}
  \label{tab:Settingup}
  \centering
  \begin{tabular}{lc}\hline\hline
    Number of SPs & 1 \\\hline
    Number of CSs & 1 \\\hline
    Number of users       & \begin{tabular}{c}
      2,048\\
      (We uniformly assigned users to\\ each effective leaf node defined by\\{\sf CompSubTree})
    \end{tabular}\\\hline
    Number of requests in each $t$     & $2^{17} = 131,072$ (each user requests 64 contents)\\\hline
    Number of contents     & 65,535 \\\hline
    Cache capacity in CS & 4,096, 8,192 and 16,384 \\
    (Maximum number of stored contents) & \\\hline
    Size of each content [MB]  & 1  \\\hline
    Popularity of content & \begin{tabular}{c}
      Zipf function in Go standard library \textbf{math/rand}\\ with arguments $s=3,v=3,000$.\\ The arguments are determined\\ by the cache hit ratio when it becomes $75\%$\\ of the cache capacity $4,096$.
    \end{tabular}\\\hline
    $T_{\sf max}$ & 16 (depth of the binary tree is 5)\\\hline\hline
  \end{tabular}
\end{table*}

\subsection{Implementing Components}

To evaluate the cache system with the mechanism described in Section~\ref{sec:tdac}, we experimentally implemented a cache system that provides time-dependent access control. The cache system is an extended version of the Cache-22 system to enable the encryption and decryption of contents with multiple keys. Three types of program code sets were implemented, namely, SP, CS, and User, which correspond to the components in Figure~\ref{cache-22-TDAC}. All modules in these components were written in the Go language using several libraries, as described in Table \ref{tab:Library}. We employed a custom Go language~\cite{GowithCIRCL} that used CIRCL~\cite{circl} patched by Cloudflare to introduce PQC primitives in addition to conventional TLS algorithms such as ECDSA and RSA. 

We implemented the SP as a web server which received requests from users to obtain ${\sf tag}_{i,j}$ via $({\sf c\_name}, j)$ as illustrated in Figure~\ref{cache-22-TDAC}. 
We also implemented the CS as a web server to forward user requests to the SP or to return cached encrypted contents to users according to ${\sf tag}_{i,j}$. 
User was a simulation program to emulate many users to get encrypted contents from the CS and decrypting them when they had the corresponding decryption key. 
Although users send requests for various contents, the popularity follows a characteristic trend, such as Zipf's law and gamma distribution, especially in the case of video-on-demand services~\cite{Cheng2013}. 
 Although all components were parameterized to adapt to various situations, we set up the experimental conditions as presented in Table~\ref{tab:Settingup} for reasonable discussion.

As the underlying cache system, we employed the Least Recently Used (LRU) cache system. That is, ciphertexts generated in the past were unavailable at the current time and were erased from the cache table ${\sf CacheTbl}$. 


\begin{figure*}[t]
\centering
\includegraphics[width=1.0\linewidth]{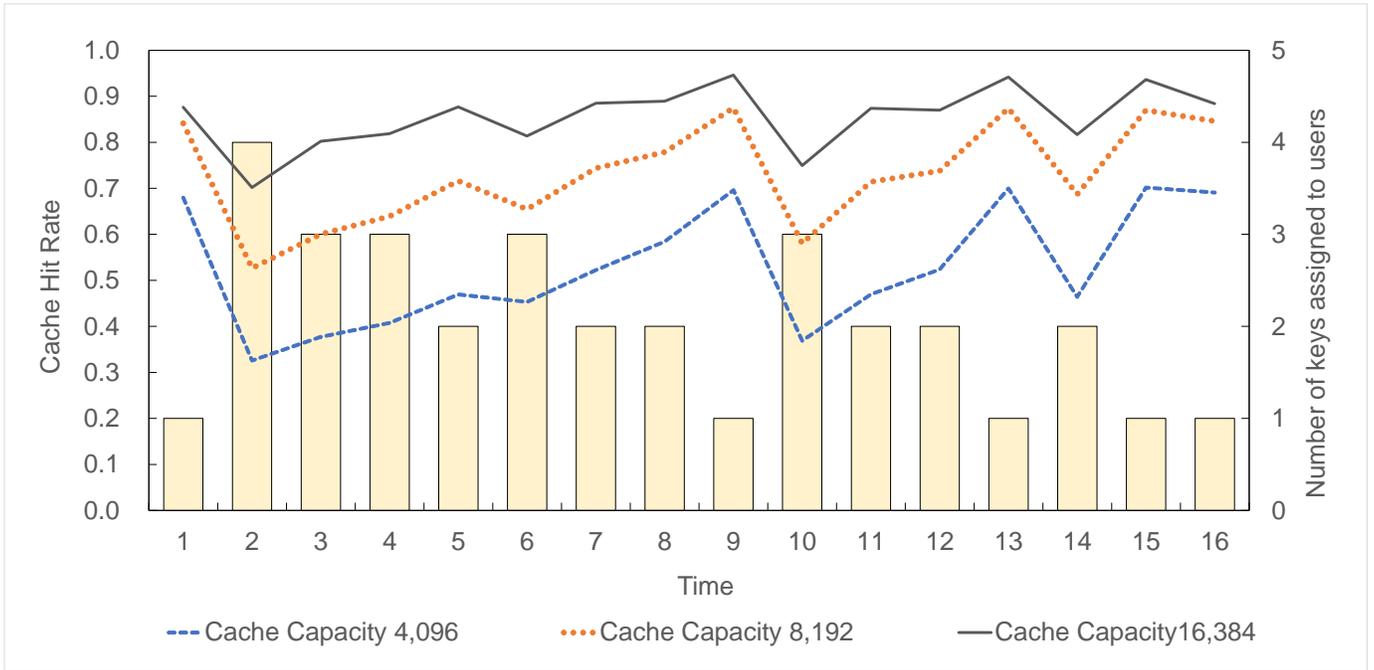}
\caption{Time Series of Cache Hit Ratio for Three Cache Capacities in CS}
\label{fig:cachehitratio}
\end{figure*}

We set up several virtual machines on Amazon Elastic Compute Cloud (EC2) with a uniform configuration, as displayed in Table~\ref{tab:HostConfiguration}. Each host ran SP and CS processes. 
Figure~\ref{fig:experimental} indicates which nodes were activated at each time period. 
Many user processes were also run on EC2 with the same configuration to emulate multiple users sending requests to obtain contents from the CS according to the sequence illustrated in Figure~\ref{fig:experimental}. Each green node indicates which key associated with the node was used for encryption. 

We set $T_{\sf max}=16$. First, we check whether no revoked users can obtain a content when users directly communicate with SP (i.e., no cache system is introduced). We emphasize that a user is revoked means that the time period corresponding to the user's access rights has passed. Next, we check that, by introducing CS, we can reduce communications between CS and SP without detracting time-dependent access control. 

\subsection{Change in Network Traffic by Introducing Time-Dependent Access Control}

A cache system is helpful to reduce traffic in a more upstream network, such as that between the CS and SP. There were two evaluation perspectives: (i) reduction in network traffic due to the cache system and (ii) increase in network traffic due to the time-dependent access control protocol. 
Figure~\ref{fig:cachehitratio} presents the time series of the cache hit ratio for each cache capacity. The three lines demonstrate that the cache capacity explicitly contributed to the reduction in network traffic. The condition of the popularity distribution in the experiment is presented in Table~\ref{tab:Settingup}. 

At $t_{1}$, all users had $k_1$ (which was assigned to the root node) and could obtain all contents encrypted by $k_1$. 
This signifies that a user could always decrypt a ciphertext that was stored due to a previous request by another user. The cache hit ratio in this situation was that same as that in a cache system without time-dependent access control. The cache hit ratio was greater than $70\%$ in all cases, 
which demonstrates that the network traffic was reduced due to the cache system. 
The reduction in network traffic was approximately $50\%$ when the cache capacity was $4,096$ MB (since the size of each content is 1 MB in our experiment) which contained $6.25\%$ of all contents. 
It could be increased to over $70\%$ when the cache capacity was increased, such as to $8,192$ MB and $16,384$ MB, which contained $12.5\%$ and $25\%$ of all contents, respectively. This indicates that the network traffic can be further reduced when time-dependent access control is employed. 

Next, we discuss how the cache capacity affects the hit ratio when employing time-dependent access control. Due to time-dependent access control, for every content, multiple encrypted data are generated with different encryption keys. 
The number of keys assigned to each content increases the number of duplicated contents. This situation may reduce the cache hit ratio because a user may not be able to decrypt a ciphertext that was stored due to a previous request by another user. The cache hit ratio is increased when the probability that the corresponding ciphertext is stored on the CS increases. Thus, when a relatively large number of keys are used for encryption, the low cache capacity of the CS may cause an increase in the cache miss rate, which increases the amount of traffic. 
The cache capacity represents the effectiveness when employing time-dependent access control. 
For example, one key is used at $t_1$, $t_9$, $t_{13}$, $t_{15}$, and $t_{16}$ (i.e., there is one green node in Figure~\ref{fig:experimental} at these time periods). Then, the differences in the hit ratio between the three cache capacities are relatively small. In contrast, four keys are used at $t_2$, and then the differences are relatively large. 
This prompts us to carefully select $T_{\sf max}$ because it depends on the depth of the binary tree and the number of keys used for encryption, although it provides more fine-grained access control. 

\section{Conclusion}

In this paper, we add a time-dependent access control protocol to the Cache-22 system and provide experimental results. Due to the proposed time-dependent access control, the number of duplicated contents is higher than that in the original Cache-22 system. That is, the proposed protocol is not only effective for controlling access rights, but it also affects the relationship between the cache capacity and network traffic. 

The prototype implementation of the original Cache-22 system considered multiple CSs and employed the color-based cooperative cache system~\cite{NakajimaYWY17}, which associates servers and caches through a color tag. 
In the Cache-22 system with time-dependent access control, a key associated with a higher node (i.e., a node closer to the root) is assigned to more users than a key associated with a lower node (i.e., a node closer to a leaf). That is, it should be effective to introduce multiple CSs that store ciphertexts encrypted by keys associated with a higher node. Confirming the effectiveness of introducing multiple CSs is left for future work. 

\medskip
\noindent\textbf{Acknowledgment}: We thank the reviewers of ICISSP 2023 for their invaluable comments and suggestions. 
This work was supported by JSPS KAKENHI Grant Number JP21K11897. 


\end{document}